\documentclass[prl,showpacs,twocolumn,preprintnumbers,amsmath,amssymb]{revtex4}

\usepackage{graphicx}

\begin{document}
%\begin{flushright}
%\end{flushright}
\title{ \bf   Characterizing symmetry breaking patterns in a lattice by dual degrees of freedom }
\author{  Yan Chen$^{1}$ and Jinwu Ye $^{2,3}$  }
\affiliation{
$^{1}$Department of Physics, State Key Laboratory of Surface Physics and Lab of Advanced Materials, Fudan
University, Shanghai, 200433, China\\
$^{2}$ Department of Physics and Astronomy, Mississippi State
University, P. O. Box 5167, Mississippi State, MS, 39762   \\
$^{3}$ Department of Physics, Capital Normal University,
Beijing, 100048 China }
\date{\today}

\begin{abstract}
    A duality transformation in quantum field theory is usually established first through partition
    functions. It is always important to explore the dual relations between various correlation
    functions in the transformation. Here, we explore such a dual relation to study quantum phases and phase transitions in an  extended boson Hubbard model
    at  $1/3$ ($2/3$) filling on a triangular lattice.
    We develop systematically a simple and effective way to use the vortex degree of freedoms on dual lattices to characterize
    both the density wave and valence bond symmetry breaking patterns of the boson insulating states in the direct lattices.
    In addition to a checkerboard charge density wave (X-CDW) and a stripe CDW, we find a novel CDW-VBS phase which has
    both local CDW and local valence bond solid (VBS) orders. Implications on QMC simulations are addressed.
    The possible experimental realizations of cold atoms loaded on optical lattices are discussed.

%    with both nearest neighbor and next nearest neighbor interactions are given.
%    We also use the method to review some known phases in the bipartite lattices.
%  So Boson Hubbard model in honeycomb lattice may show different physics than that in square lattice
\end{abstract}


\maketitle

   Various duality transformations  play crucial roles
   in non-perturbative calculations of quantum field theories, statistical mechanics and string theory.
   For example, the $ 2d $ Ising model is self-dual, the high temperature region of a $ 2d $ Ising model
   in a lattice can be mapped to the low temperature region of its dual Ising model defined on a dual
   lattice and vice versa \cite{dualrev}.
   The two spin correlation function of the Ising model can also be calculated in terms of its dual Ising model with suitably chosen frustrated bonds.
%    The 3d Ising model is dual to a 3d $ Z_2 $ lattice gauge theory \cite{dualrev}.
   The high temperature region of a 3d Ising model  can be mapped to the low temperature
   region of a 3d $ Z_2 $ lattice gauge theory defined on the links of its dual lattice.
%   The two spin correlation functions of the 3d Ising model
%   can also be evaluated in terms of the 3d $ Z_2 $ lattice gauge
%   theory in its dual lattice with some frustrated plaquettes. While the correlation functions of a Wilson loop in the 3d $ Z_2 $ lattice gauge
%   theory in its dual lattice can also be evaluated in terms of the 3d Ising
%   model with some frustrated bonds\cite{dualrev}.
   It was also shown in \cite{3dxy} that the high temperature region of a 3d $ XY $ model is dual to the low temperature region of a 3d
   $ U(1) $ lattice gauge field defined on its dual lattice.
%    The duality between one field theory and its corresponding dual field theory is usually
%    established through the partition functions \cite{dualrev}.
%    For example, the $ 2d $ Ising model is self-dual, the
%    3d Ising model is dual to a 3d $ Z_2 $ lattice gauge theory, 2d
%    $ XY $ model is dual to a $ 2d $ Coulomb gas model, etc.
%    Especially, the 3d $ XY $ model in a lattice is dual to a Abelian $ U(1) $ lattice gauge theory
%    in a dual lattice \cite{3dxy}. In the context of the boson Hubbard model Eqn.\ref{boson},
%    this duality shows that interacting bosons at integer fillings $ f=n $ hopping on
%    a lattice is dual to interacting vortices hopping on a dual
%    lattice in the presence of fluctuating non-compact $ U(1)$ gauge field.
%    However, the duality relations between various correlation
%    functions in the two theories are much more involved. These correlation functions
%    could be the original order parameter, their composite operators
%    such as a current, density or spin operator or a Wilson loop.
%    One of the authors  \cite{gauge1,gauge2} discussed the importations of the  gauge invariant electron Green functions
%    in the pseudo-gap regime of the  high temperature superconductors and its connections to the Angle-Resolved Photo-Emission  Spectroscopy (ARPES).
%    Various S-dualities interconnect different string theories and plays important roles to .
%    In principle, the correlation functions of the original variables in a direct lattice can
%    be calculated in terms of the dual variables in the dual lattice.
     In the recently discovered $ AdS_{d+1}/CFT_{d} $ correspondence,
%     \cite{adscft} established a duality between the conformal field theory at $ d $
%    dimension  ( $ CFT_{d} $ ) in the boundary and the gravity at one higher dimension (
%    Holographic principle ) in anti-de Sitter space ( $ AdS_{d+1} $ ) in the bulk.
     several correlation functions of the  strongly coupled quantum field theory in the flat space on its boundary can be
     computed just by solving the equation of motions in a classical gravity in the asymptotic $ AdS $ geometry in  the bulk \cite{adscftapply}.

     Recently,  by a dual vortex method (DVM), Balents \emph{et al.}~\cite{pq1} studied the quantum phases and phase transitions of the extended boson Hubbard model (EBHM)
     in a square lattice at generic commensurate filling factors $f=p/q$ ( $p, q$ are relative prime numbers ).
%     systematically studied in \cite{pq1}.
%     After performing a charge-vortex duality transformation similar to the one used in the 3d $ XY $ model \cite{3dxy},
     They mapped the interacting bosons at the filling factors $f$ hopping in a lattice
     to the interacting vortices  hopping on the dual lattice subject to a fluctuating
     {\em dual} magnetic field whose average strength  through a dual plaquette is equal to the boson density $f=p/q$.
    The projective representation of the space group (PSG) dictates that there are at least $ q $-fold degenerate minima  $\phi_{l}, l=0,1,\cdots, q-1$
    in the mean field energy spectrum. In the continuum limit, the effective action describing the superfluid (SF) to an insulator transition in terms of
    these $q$ order parameters should be invariant under this PSG. They also constructed a density operator formalism (DOF) to
     describe the symmetry breaking patterns of bosons in the direct lattice.
     Later, the DVM was applied to a triangular lattice \cite{tri}.
     The DVM method was applied by one of the authors \cite{five,univ} to study quantum phases and phase transitions
     of the EBHM in a honeycomb lattice at and near half fillings. Despite all these previous studies, one remaining
     outstanding problem is how to characterize the symmetry breaking patterns of bosons in a insulating state in terms of the dual
     vortices in the corresponding dual lattice.
%    The duality mapping of Eqn.\ref{boson} is not complete without    resolving this outstanding problem.
     Here we attempt to resolve this outstanding problem. 
     
    We develop a systematic way to determine the symmetry breaking patterns of the insulating states
    in terms of the vortex degree of freedoms only at the dual honeycomb lattice points.
    These vortex degree of freedoms are  the gauge invariant physical vortex densities,
    the kinetic energies and vortex currents defined in Eqn.\ref{density} and \ref{bond}.
    We find the checkerboard charge density wave (X-CDW) phase in Fig.\ref{xstripe}a in the Ising limit,
    the stripe phase in  Fig.\ref{xstripe}b in one of the easy plane limit, both of which were
    found previously by the DOF \cite{tri}. Most importantly, we identify a novel CDW-VBS phase which has
    both CDW and valence bond solid (VBS) orders shown in Fig.\ref{cdwvb} in another easy plane limit.
    This CDW-VBS phase differs from the bubble solid phase  in the same easy plane limit found by the DOF
    in \cite{tri}. This disagreement shows that the  DOF developed in \cite{pq1,tri} is at least incomplete.
    The method developed here should be very general and can be used to characterize uniquely the symmetry
    breaking patterns of bosons in any lattices at any  filling factors
    by using the vortex degree of freedoms only at the corresponding dual lattice points. 
    We compare our results with the previous Quantum Monte Carlo (QMC) results in a triangular lattice \cite{trinnn}
    with $ V_{1} $ and  $ V_{2} $ interactions and also give important implications
    to possible future QMC simulations. The quantum phases identified in this paper may be realized in the
    future experiments of the dipolar bosons loaded on a triangular lattice \cite{junpolar}.

\begin{figure}
\includegraphics[width=6.0cm]{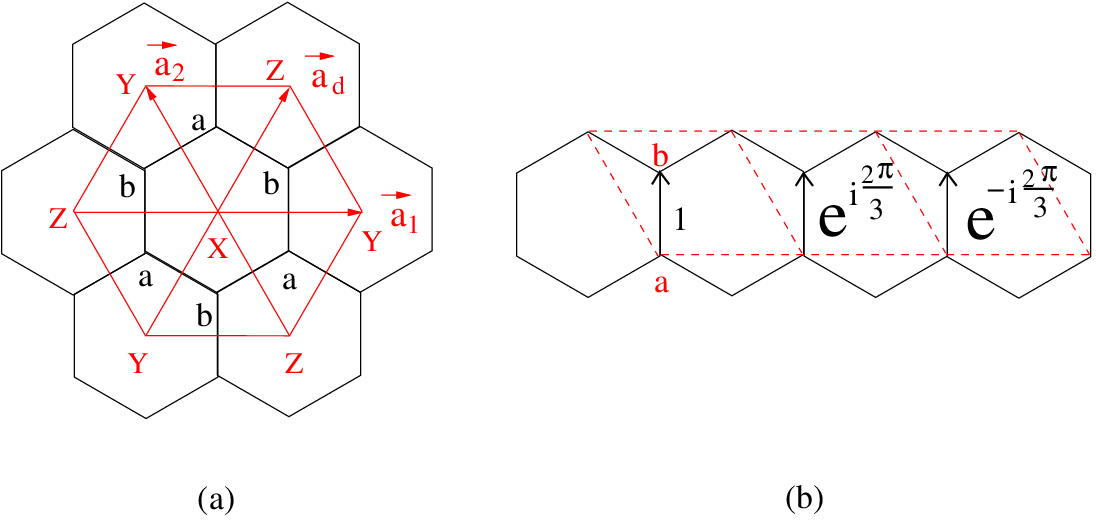}
\caption{(Color online). (a) Bosons at filling factor $ f $ are
hopping on a triangular lattice ( red line ) which has three
sublattices $ X, Y, Z $. Its dual lattice is a honeycomb lattice (
black line ) which has two sublattices $ a $ and $ b $. (b) The bond
phase factors in a dual honeycomb lattice at $ f=1/3 $ flux quanta
per hexagon. The direction of the gauge field is important. At $
f=2/3 $, the bond phase factors are just complex  conjugate of those
at $ f=1/3 $. } \label{trigauge}
\end{figure}

%     In a triangular lattice in Fig.1a at the filling factor $ f=1/3 $ with

      The EBHM  in both bi-partisan and  frustrated lattices with various  kinds of interactions and filling factors
     (commensurate or in-commensurate) is described by the following Hamiltonian \cite{pq1,gan,five}:
\begin{eqnarray}
  H  & = & -t \sum_{ \langle ij \rangle } ( b^{\dagger}_{i} b_{j} + h.c. ) - \mu \sum_{i} n_{i} +
       \frac{U}{2} \sum_{i} n_{i} ( n_{i} -1 )     \nonumber  \\
      & + &  V_{1} \sum_{ \langle ij \rangle } n_{i} n_{j}  + V_{2} \sum_{ \langle\langle ik \rangle\rangle } n_{i} n_{k} + \cdots
\label{boson}
\end{eqnarray}
    where $ n_{i} = b^{\dagger}_{i} b_{i} $ is the boson density, $
    t $ is the hopping amplitude,
    $ U, V_{1}, V_{2} $ are onsite, nearest neighbor and next nearest neighbor interactions
    between the bosons. The $ \cdots $ may include further neighbor interactions, the dipole-dipole interaction \cite{dipolarss,dipolarsstri},
    3-body interaction \cite{3body}  and possible ring-exchange interactions \cite{ring}.
    The EBHM Eqn.\ref{boson} can be realized by ultra-cold atoms
    loaded in optical lattices \cite{manybody}.  For example, using three coplanar beams of equal intensity having
    the three vectors making a $ 120 ^{\circ}$ angle with each other,
    the potential wells have their minima in a triangular lattice \cite{trilattice}.
    The long range interactions $ V_1, V_2, \cdots $ and the dipole-dipole interaction can be realized with dipolar bosons loaded in optical lattices \cite{junpolar}.
    In this paper, we will use the DVM developed in \cite{pq1,univ}
    to study various kinds of insulating phases and phase transitions in a triangular lattice
    (Fig.\ref{trigauge}a) at  $ 1/3 (2/3) $ fillings.
     With the gauge chosen in Fig.1b, the general effective vortex action in terms of the three modes $ \xi_{l}, l=0,1,2 $ ( See Eqn.\ref{fields} below )
     invariant  under all the PSG transformations upto sixth order terms was written down in \cite{tri}.
     In the permutative  representation basis $ \phi_{l}, l=0,1,2 $ given by $
      \phi_{0}  =  \frac{1}{\sqrt{3}} ( \xi_0 + e^{-i \frac{2 \pi}{3}} \xi_{1} + \xi_{2}
      ), \phi_{1}  =  \frac{1}{\sqrt{3}} (e^{-i \frac{2 \pi}{3}} \xi_0 +  \xi_{1} + \xi_{2}
      ),  \phi_{2}  =  \frac{1}{\sqrt{3}} ( \xi_0 +  \xi_{1} + e^{-i \frac{2 \pi}{3}} \xi_{2}  )  $,
%      Moving {\em slightly} away from the $ 1/3 $  filling $ f=1/3 $ corresponds to adding \cite{univ} a small {\em mean} dual
%     magnetic field $ \delta f= f-1/3 $  in the action derived in \cite{tri}.
     the effective vortex action  $ {\cal L}_{SF} ={\cal L}_{0} + {\cal L}_{1} + {\cal L}_{2} $ can be simplified   to:
\begin{eqnarray}
    {\cal L}_{0} & = &  \sum_{l} | (  \partial_{\mu} - i A_{\mu} ) \phi_{l} |^{2} + r | \phi_{l} |^{2}
    +  ( \epsilon_{\mu \nu \lambda} \partial_{\nu} A_{\lambda} )^{2}/4
       \nonumber  \\
    {\cal L}_{1} &  =  & u ( | \phi_{0} |^{2} + | \phi_{1} |^{2}+ | \phi_{2} |^{2}  )^{2}
       - v [ ( | \phi_{0} |^{2} - |\phi_{1} |^{2} )^{2}
                        \nonumber    \\
          & + & ( | \phi_{1} |^{2} - |\phi_{2} |^{2} )^{2}
          +  ( | \phi_{2} |^{2} - |\phi_{0} |^{2} )^{2} ]
                                   \nonumber  \\
     {\cal L}_{2} & = &  w [ (\phi^{*}_{0} \phi_{1} )^{3}+ (\phi^{*}_{1} \phi_{2}
           )^{3}+ (\phi^{*}_{2} \phi_{0} )^{3} +h.c.]
\label{tri}
\end{eqnarray}
      where $ A_{\mu} $ is a non-compact  $ U(1) $ gauge field. When $ r > 0 $, the bosons are in the superfluid state,
      $  \langle \phi_{l} \rangle =0 $ for every $l=0,1,2$. When $ r <  0 $, the bosons are in a insulating state,
      $ \langle \phi_{l} \rangle \neq 0 $ for  at least one $ l $.

    In order to develop a systematic way to use the vortex degree of
    freedoms in the honeycomb lattice to describe the symmetry
    breaking patters of the bosons in the triangular lattice, one
    has to derive the relation \cite{dof} (namely Eqn.\ref{fields} below) between the total vortex fields
    in the honeycomb lattice and the order parameters $ \phi_{l} $
    in Eqn.\ref{tri}. Then one need to first study \cite{five} the energy spectrum of
    the vortices hopping in the honeycomb lattice in the presence of  $ f=p/q $ flux quantum per hexagon shown in Fig.1a.
    For the gauge chosen in Fig.1b, the vortex hopping Hamiltonian is $ H_{v}=- t_{v} \sum_{\vec{x}}[ |\vec{x}+ \vec{\delta}
    \rangle e^{i 2 \pi f a_{1} } \langle \vec{x} | + |\vec{x}+ \vec{\delta} \rangle \langle
    \vec{x} + \vec{a}_{2} | + |\vec{x}+ \vec{\delta} \rangle \langle \vec{x} +
    \vec{a}_{1}+ \vec{a}_{2} | +h.c. ] $ where  $ \vec{x}= a_1 \vec{a}_1 + a_2 \vec{a}_2 $
    and $ \vec{x} + \vec{\delta },  \vec{\delta }= 1/3 \vec{a}_1 + 2/3 \vec{a}_2 $  belong to the sublattice $a$ and $b$ respectively in Fig.1b.
    Following the Sec.3 of Ref.\cite{five}, one can derive the Harper's equation corresponding to $ H_v $:
\begin{eqnarray}
    c^{a}_{m-1}( k_x,k_y )  & +  & e^{i k_y}( 1+ e^{i ( k_x + 2 \pi f m ) } ) c^{a}_{m}( k_x,k_y)     \nonumber  \\
                            &  = &  \epsilon( k_x,k_y)   c^{b}_{m}( k_x,k_y),   \nonumber  \\
     c^{b}_{m+1}( k_x,k_y ) & +  & e^{-i k_y}( 1+ e^{-i ( k_x + 2 \pi f m ) } ) c^{b}_{m}( k_x,k_y)   \nonumber  \\
                             & = & \epsilon( k_x,k_y)   c^{a}_{m}( k_x,k_y)
\label{harper}
\end{eqnarray}
   where $ m=0,1,\cdots,q-1 $ and $ a,b $ are two sublattices of the honeycomb
   lattice, the  $ - \pi/q \leq k_x \leq \pi/q,  - \pi \leq k_y \leq \pi $ are inside the reduced Brillouin zone.
%   The $ q $ minima of the $ 2 q $ bands $ \epsilon( k_x,k_y) $ was found to be at $ (  k_x,k_y )=(0, 2\pi f l ), l=0,1,\cdots,q-1 $.
   At $ q=3 $, we are able to find the analytic expressions of the  eigenvalues and the
   corresponding eigenvectors of the $ 6 \times 6 $ matrix in Eqn.\ref{harper} at the 3 minima  $ (  k_x,k_y )=(0, 2\pi l/3 ), l=0,1,2 $.
%   One also need to find the eigenvector
%   $ c^{a}_{m} (l=0)= c^{a}_{m}, c^{b}_{m} (l=0)= c^{b}_{m} $ at $ l=0 $.
    The lowest eigenvalue is $ \epsilon= -( 1+ 2\cos 2 \pi/9 ) t_v $, the corresponding eigenvector at $ l=0 $ is
    $  (c^{a}_{m}(l=0), c^{b}_{m}(l=0) )= [ 2\cos 4\pi/9 + 2\cos 2\pi/9 +1, e^{-i \pi/3}, 2 \cos 2\pi/9,
       2 \cos 4\pi/9 + 2 \cos 2\pi/9 +1, 2\cos 2\pi/9, e^{-i \pi/3}  ],~~ m=0,1,2 $.
%    Note that $ c^{b}_{0}(l=0)= c^{a}_{0}(l=0), c^{b}_{1}(l=0)=c^{a}_{2}(l=0), c^{b}_{2}(l=0)=c^{a}_{1}(l=0) $.
    The eigenvectors at $ l=1,2 $ can be achieved by
    the magnetic translation \cite{pq1,tri} along $ \vec{a}_{1} $: $ c^{a}_{m} (l) =c^{a}_{m}(l=0) \omega^{-ml} , c^{b}_{m} (l)=  c^{b}_{m}(l=0)\omega^{-ml} \omega ^{l}  $
    where $ \omega = e^{i 2 \pi f } $.
    The eigenfunctions at the three minima $ (  k_x,k_y )=( 0, 2\pi l/3 ) $ are
    $  \psi^{a}_{l}( \vec{x} )= \sum^{q-1}_{m=0} c^{a}_{m} (l) e^{i 2 \pi f ( m a_1+ l  a_2 ) },
    \psi^{b}_{l}( \vec{x} )= \sum^{q-1}_{m=0} c^{b}_{m} (l) e^{i 2 \pi f ( m a_1+ l a_2 )} $.
%    One can write the two component vortex field at  $ (  k_x,k_y )=( 0, 2\pi l/3 ) $ as
%    $ \Psi_{l}( \vec{x} )= ( \psi^{a}_{l}( \vec{x} ), \psi^{b}_{l}( \vec{x} ) ) $ .
    Then one can write the {\bf total} vortex fields at the sublattice a and b in the Fig.1 as  the expansion in terms of the three eigenfunctions:
\begin{equation}
    ( \Phi^{a}( \vec{x} ), \Phi^{b} ( \vec{x} ) )^{T} = \sum^{2}_{l=0}  ( \psi^{a}_{l}( \vec{x} ), \psi^{b}_{l}( \vec{x} ) )^{T} \xi_{l}
\label{fields}
\end{equation}
     where the $ T $ means the transpose and the three coefficients $\xi_{l}, l=0,1,2 $ are the three vortex order parameters.
     The effective action Eqn.\ref{tri} was written in the permutative representation $ \phi_{l} $.
% listed above Eqn.\ref{tri}.

    Plugging the mean field solutions of all insulating states where $ r < 0 $ in Eqn.\ref{tri} into Eqn.\ref{fields},
    one can construct the following gauge invariant quantities \cite{hightc}:
    the densities at different sites in sublattices $ a $ and $ b $:
\begin{equation}
   | \Phi_{a}( \vec{x} ) |^{2}, ~~~~~| \Phi_{b}( \vec{x} ) |^{2}
\label{density}
\end{equation}
     and the bond quantities between sublattice $ a $ and sublattice $ b $ :
\begin{equation}
    \Phi^{\dagger}_{b}( \vec{x} ) e^{i 2 \pi f a_1} \Phi_{a}( \vec{x} ) = K - i I
\label{bond}
\end{equation}
    where $ \vec{x} $ belongs to the same unit cell shown in Fig.1b. All the other bonds having no phase factor from the gauge field.
    The real part $ K $ gives the kinetic energy between the two sites. The imaginary part $ I $  gives the current between the two sites.
%    Note that the minus sign in Eqn.\ref{bond} is important which make the definition of the current to be consistent with
%    that of gauge factors in the Fig.1b.
    By using the Eqn.\ref{density}, especially Eqn.\ref{bond}, one can  extract the corresponding symmetry breaking patterns of bosons in the direct triangular lattice.
    In the following, we discuss the 3 possible insulating states of  Eqn.\ref{tri} where $ r < 0 $ respectively.

%     If $ r > 0 $, the system is in the superfluid state $  < \psi_{l} > =0 $ for every $ l $.
%     If $ r < 0 $, the system is in the insulating state $ < \psi_{l} > \neq 0 $ for
%     at least one $ l $.

 (1) If $ v>0 $, the system is in the Ising limit, only one of the 3
 vortex fields condense. For example, substituting $ \phi_0=1, \phi_1=\phi_2=0 $ into
 Eqns.\ref{fields}-\ref{bond}, one can evaluate the vortex densities, kinetic energy and the current in the dual honeycomb lattice.
 For simplicity, we only show the currents in Fig.\ref{xstripe}a which are obviously conserved.
 It is important to observe that the chirality  $  \chi_{p}= \sum_{p} I $ around any hexagon $ p $  leads to the boson density at the center of
 the hexagon $ n_p = 1/3 + \chi_{p} $.
% By counting the segments of currents along the bonds surrounding the $X, Y, Z $ lattice points, paying special attentions to
%   their counter-clockwise or clockwise directions,
   Then we can calculate  the densities at the three sublattices: $ n_{x}=1/3 + 6 \delta I > 1/3, n_{y}=n_{z}= 1/3 - 3 \delta I < 1/3
   $ with the constraint $ n_x + n_y + n_z= 1 $ where the $ I $ is the current flowing around the $ X $ lattice site in Fig.\ref{xstripe}a.
   The $ \delta \sim n_{x}-n_{y} >0 $ can be taken as the CDW order parameter measuring the distance from the SF to the CDW transition.
%   When one tunes the density $ n_{y}=n_{z}=0 $ which stand for vacancies, then $ n_{x}= 1 $ which stands for one boson.
%   This corresponds to the classical limit $ t=0 $ in Eqn.\ref{boson}. With the quantum fluctuations $ t >0 $, then $ n_{x} <1, n_{y}=n_{z} > 0 $.

\begin{figure}
\includegraphics[width=3.0cm]{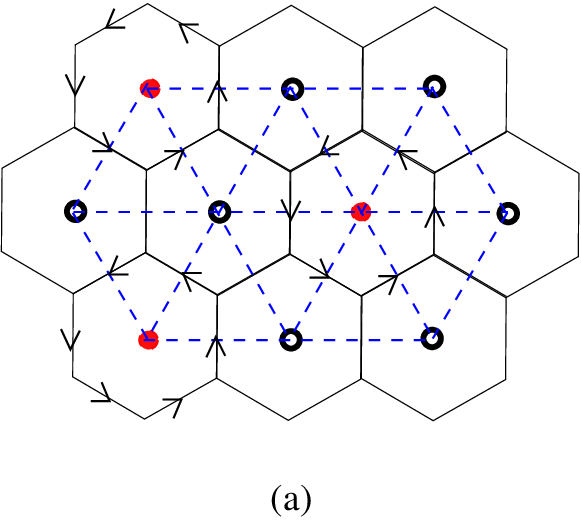}
\hspace{0.5cm}
\includegraphics[width=3.0cm]{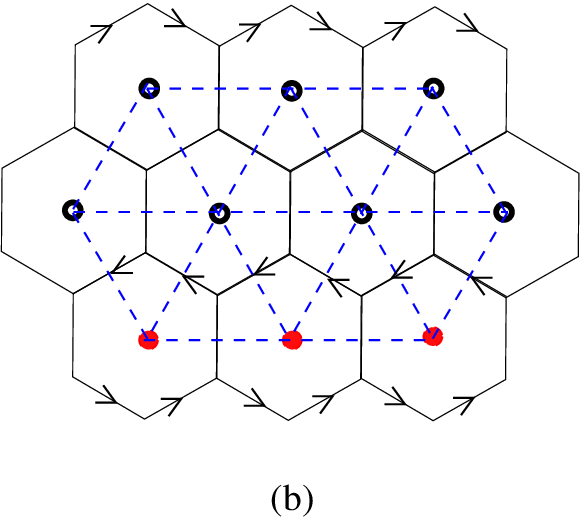}
\caption{(Color online). (a) The X-CDW state in the Ising limit $ v
> 0  $ at $ f=1/3 $.
 (b) The period 3 stripe phase in the easy plane limit $
v<0, w< 0 $ at $ f=1/3 $.  At $ f=2/3$, one can just reverse the
current flow and perform a particle and hole
 transformation on $ f=1/3 $, namely, exchange particles ( red dots )  and vacancies ( black empty circles ).  }
\label{xstripe}
\end{figure}

 (2) If $ v<0 $, the system is in the easy-plane
 limit, all the three vortex fields have equal magnitude $ \phi_{l}= |\phi| e^{i \theta_{l}} $ and
 condense. There are also two distinguished  cases:

 (2a) If $ v<0, w < 0 $, the mean field solution \cite{tri} is $ \theta_1-\theta_0=  2 \pi m/3, \theta_2 - \theta_0= 2 \pi n/3  $ where $ m, n=0,1,2 $.
 Substituting $ \theta_0= \theta_1=\theta_2=0 $ into Eqns.\ref{fields}-\ref{bond},
 one can evaluate the vortex densities, kinetic energy and the current in the dual honeycomb lattice.
 For simplicity, we only show the currents in Fig.\ref{xstripe}b. Similar arguments as in (1) shows that this is a stripe phase.

 (2b)  If $v >0,  w>0 $, one of the 18 equivalent solutions \cite{tri} is $
\theta_1-\theta_0= \theta_0 - \theta_2= 2 \pi/9 $, then $ \theta_1-
\theta_2= 4 \pi/9 $.  Substituting this solution into Eqns.\ref{fields}-\ref{bond},
   one can evaluate the vortex densities, kinetic energy and the currents shown in
   Fig.\ref{cdwvb}. It is important to realize there are two
   independent currents $I_1$ and $I_2$ flowing in Fig.\ref{cdwvb}.
%   By counting the number of currents along the bonds surrounding the red, green and black lattice points, paying special attentions to
%   their counter-clockwise or clockwise directions,
   Similar to (1), we can calculate the three different densities: $ n_{r}=1/3 + 2 \delta [ I_{1} + ( I_{1}- I_{2} ) ] > 1/3,
   n_{g}=1/3 + 2 \delta [ I_{2} + ( I_{2}- I_{1} ) ] > 1/3,
   n_{b}=1/3-2 \delta [ I_{1}+ I_{2} ]  < 1/3 $ with the constraint $ n_r+ n_g + n_b =1 $
   where $ I_{1}= \sin \frac{ 3 \pi}{9} +  \sin \frac{ 2 \pi}{9} > I_{2}= \sin \frac{ 2 \pi}{9} +  \sin \frac{ \pi}{9} $.
   The $ \delta $ can be taken as the CDW-VBS order parameter measuring the distance from the SF to the CDW+VBS transition.
   When one tunes the density $ n_{b}=0 $, then $ n_{r}= \frac{ I_{1} }{ I_{1} + I_{2} } > n_{g} = \frac{ I_{2} }{ I_{1} + I_{2} } $
   with $ n_{r}/n_{g} = I_{1}/I_{2}, n_{r}+ n_{g}=1 $.

%   At $ f=2/3 $ case, one only need to change $ n_{r} \rightarrow  \tilde{n}_{r} = 1-n_{r}=2/3 - 2 \delta [ I_{1} + ( I_{1}- I_{2} ) ] < 2/3,
%   n_{g} \rightarrow \tilde{n}_{g}= 1-n_{g}=2/3 - 2 \delta [ I_{2} + ( I_{2}- I_{1} ) ] < 2/3,
%   n_{b} \rightarrow \tilde{n}_{b}= 1-n_{b}=2/3 + 2 \delta [ I_{1}+ I_{2} ] > 2/3 $
%   withe constraint $ \tilde{n}_r+ \tilde{n}_g + \tilde{n}_b =2 $.
%   When one tunes the density $ n_{b}=1 $, the $ n_r=\frac{ I_{2} }{ I_{1} + I_{2} } <  n_{g}= \frac{ I_{1} }{ I_{1} + I_{2} } $ with $ n_{r}+ n_{g}=1 $.
%   When one tunes the density $ \tilde{n}_{r}=0 $, then
%   $\tilde{n}_{g}= \frac{ 2 (I_{1} -I_2 ) }{  2I_{1} - I_{2} },  \tilde{ n}_{b} = \frac{ 2 I_{1} }{  2I_{1} - I_{2} } $ with $ \tilde{n}_{g} + \tilde{n}_{b}=2 $.

\begin{figure}
\includegraphics[width=6.0cm]{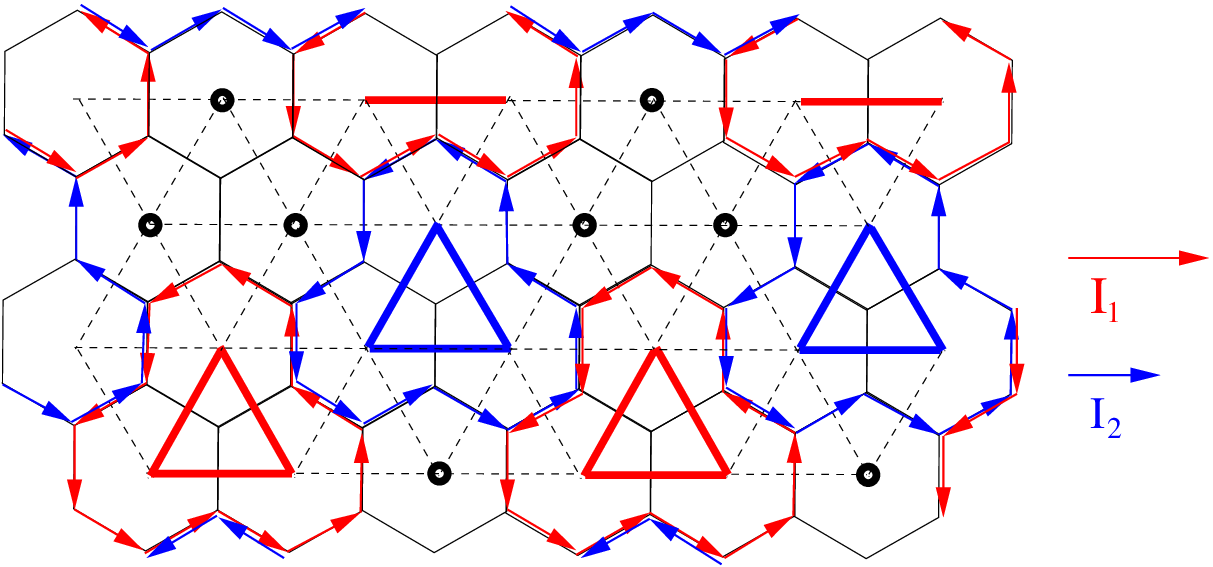}
\caption{(Color online). The CDW-VBS phase in a triangular lattice
at $ f=1/3 $ in the easy plane limit $ v<0, w> 0 $. Shown is $
\theta_0=0, \theta_1= 2 \pi/9, \theta_2= -2 \pi/9 $ case. There are
two different vortex currents flowing in the dual honeycomb lattice.
The red current is $ I_{1}= \sin \frac{ 3 \pi}{9} + \sin \frac{ 2
\pi}{9} $, the green current $ I_{2}= \sin \frac{ \pi}{9} + \sin
\frac{ 2 \pi}{9} $. The vortex currents are conserved at all the lattice
points. The two different vortex currents in a dual honeycomb lattice imply
3 different densities $ n_r,n_g,n_b $ in a triangular lattice listed
in the text. At $ f=2/3$, one can just reverse the current flow and
perform a particle and hole transformation on $ f=1/3 $.}
% For $ \theta_0=0, \theta_1=
%-2\pi/9, \theta_2= 2 \pi/9 $,
% one only need to reflect the figure with respect to the $ \vec{a}_{1} $. }
\label{cdwvb}
\end{figure}

   It is important to observe that the vortex
   fields vanish  at the centers of  both the red loop and the green loop, therefore, both the kinetic energy and the current emanating from
   the centers vanish.  This fact indicates that there are local SF (or VBS order) around the
   these two dual lattice points as shown by the red and green triangles in Fig.\ref{cdwvb}.  It is interesting to compare this fact with the
   known result  that the interacting bosons in a Kagome lattice at $f=1/2$ \cite{five,ka1} are always in a SF state  due to the localization of the vortices (a flat vortex band).
   The crucial difference is that here there is only a local SF around the the centers of the red loop and the green loop, while there is a global
   SF across the whole lattice in the latter.  We need also stress that the vortex field is non-vanishing at the black point, so only the current
   vanishes, but the kinetic energy does not vanish, it shows that there are local CDW around the dual lattice point
   as shown by black circles in Fig.\ref{cdwvb}. So this state has both VBS and CDW order which is a hybrid state unique to a frustrated lattice.

%\vspace{4cm}

 Now we will determine quantitatively the CDW and VBS ordering patterns of the CDW-VBS  phase in Fig.3.
 The CDW component of the CDW+VBS phase is given by:
\begin{eqnarray}
 \rho_{CV}( \vec{x} )  &  =  &  1/3 + 4 \delta \sqrt{ \frac{I^{2}_1-I_{1} I_{2} + I^{2}_2
 }{3}} [ - \cos ( \frac{ 2 \pi }{3} x + \frac{ 5 \pi}{18} )   \nonumber \\
 & + & \cos ( \frac{ 2 \pi }{3} y  + \frac{ \pi}{18}
 ) + \cos ( \frac{ 2 \pi }{3} (x-y)  + \frac{ \pi}{18} )]
\label{directtricv}
\end{eqnarray}
  where $ \delta $ is the CDW order parameter and $ I_1, I_2 $ are
  the vortex currents in Fig.\ref{cdwvb}.  The VBS component of the CDW+VBS phase is given by:
\begin{eqnarray}
 B_{1}( \vec{x} ) &  =  &  B_{d}( \vec{x} )= c \delta ( I_1 + I_2 ) ( 1 + 2 \cos  \frac{ 2 \pi }{3} (x+y)
 )   \nonumber   \\
 & + & 2 c \delta \sqrt{ I^{2}_1-I_{1} I_{2} + I^{2}_2 }[  \cos ( \frac{ 2 \pi }{3} x + \frac{ 2 \pi}{9} )   \nonumber \\
 & + & \cos ( \frac{ 2 \pi }{3} y  - \frac{ 2 \pi}{9}
 ) + \cos ( \frac{ 2 \pi }{3} (x-y)  - \frac{ 2 \pi}{9} )]     \nonumber  \\
  B_{2}( \vec{x} ) &  =  & B_{1} ( \vec{x} - \vec{a}_1 )
\label{directtricvb}
\end{eqnarray}
   where the $ c $ is an overall constant which can not be determined from the PSG symmetry based on the DVM.
   When comparing with the CDW order in Eqn.\ref{directtricv}, we can see that in addition to the 3 ordering wave
   vectors $ \vec{Q}_{\alpha}= 2\pi/3(1,0), 2\pi/3(0,1), 2\pi/3(1,-1),\alpha=1,2,3
   $, there is also a new VBS ordering  wave vector $ \vec{Q}_{VBS}=2\pi/3(1,
   1) $. It is this new ordering wave vector which makes the determination of the VBS order  inside the CDW-VBS phase
   possible by the QMC \cite{trinnn} and
   its detection  possible by light scattering experiments of cold atoms loaded on optical lattices \cite{bragg}.

 By extending the DOF in \cite{pq1} for a square lattice to the triangular lattice,
 the authors in Ref.\cite{tri} identified a bubble CDW  phase in the $ v>0, w>0 $ case.
 This bubble CDW phase has only one vortex current $ I $ flowing in the dual honeycomb lattice, so
 there is just two different densities. Furthermore, there is no VBS order.
 Therefore the bubble phase is completely different from the CDW-VBS phase in Fig.\ref{cdwvb}.
 This discrepancy indicates the DOF  developed in
  \cite{pq1,tri} may have intrinsic difficulties. Some general problems associated with the DOF are examined in \cite{long}.

% The density of the bubble phase is given by:
%\begin{eqnarray}
% \rho_{B}( \vec{x} )  &  =  &  1/3 + \frac{ 4 }{\sqrt{3}} \delta I [ \cos ( \frac{ 2 \pi }{3} x -\frac{  \pi}{2} )   \nonumber \\
% & + & \cos ( \frac{ 2 \pi }{3} y  - \frac{ \pi}{6}
% ) + \cos ( \frac{ 2 \pi }{3} (x-y)  - \frac{ \pi}{6} )]
%\label{directtrib}
%\end{eqnarray}
%  Although  Eqn.\ref{directtricv} and
%  Eqn.\ref{directtrib} have the same ordering wave-vectors $   \vec{Q}_{\alpha}= 2\pi/3(1,0), 2\pi/3(0,1),
%  2\pi/3(1,-1),\alpha=1,2,3 $, the three phases inside the $ \cos $ functions are different.
%  Most importantly, there is no VBS order in the bubble phase.

%  QMC simulations can be directly performed on the EBHM and can be compared with the DVM developed in this paper.
  The  EBHM of the hard core bosons with $ U=\infty, V_{1} >  0, V_{2} >0 $
  was studied by QMC in \cite{trinnn}.  A period-3 stripe solid state in Fig.\ref{xstripe}b is found at $ f=1/3
  $.  The dual vortex effective action to describe the transition from the SF to the stripe  solid  is given by Eqn.\ref{tri}
  in the easy plane limit $ r<0, v<0, w<0 $.
%  A meta-stable bubble solid phase  was also found in the QMC in \cite{trinnn}.
%  This bubble solid phase has a higher energy than the stripe solid phase in Fig.\ref{xstripe}b.
   Unfortunately, the very interesting CDW-VBS phase in Fig.\ref{cdwvb} was not searched in the QMC in \cite{trinnn} in any parameter regimes.
  In order to identify this phase, in addition to the QMC calculations of the SF density and the density structure factor
  in \cite{trinnn}, a bond structure factor shown in Eqn.\ref{directtricvb} need also be studied to identify the VBS ordering where
  there is an additional new VBS ordering wavevector $ \vec{Q}_{VBS}= \frac{ 2 \pi}{3} (1,1) $. The authors in \cite{3body}
  studied the EBHM Eqn.\ref{boson} with a three-body interaction in a 1 dimensional lattice by QMC.
  They also found a solid state with the coexistence of CDW and VBS at $ f=2/3 $. It would be very interesting so do a QMC on a triangular lattice
  with the 3-body interaction to see if the 3-body interaction can stabilize the CDW-VBS phase in Fig.3.
%  It is also interesting to identify the corresponding  CDW-VBS-SS  slightly away from $ 1/3 $ filling for soft core bosons.

%    In summary, we used the DVM developed in \cite{pq1,univ}
%    to study various kinds of insulating phases and phase transitions in a triangular  lattice at
%    and slightly away $ 1/3 $ and $ 2/3 $ fillings.
%    At the commensurate fillings $ 1/3 (2/3) $, we developed a systematic way to identify the
%    symmetry breaking insulating states of bosons in the direct
%    lattice  uniquely and completely by using gauge invariant vortex density, kinetic energy and current on the dual lattice.
%    We reproduced the X-CDW phase in Fig.\ref{xstripe}a and the Stripe phase in  Fig.\ref{xstripe}b.
%    Most importantly, we also identify a novel CDW+VBS phase which has both CDW and VBS
%    orders and is unique to frustrated lattices.

%    In summary, we developed a systematic method to determine the
%    symmetry breaking insulating states of bosons in the direct
%    lattice  uniquely and completely by using gauge invariant vortex density, kinetic energy and current on the dual lattice.
%    The method developed in this paper is very general and  can be applied to any lattices such as bipartite lattices and frustrated lattices.
  
    When our method is applied to bipartite lattices \cite{long}, it can recover all the previously known phases in \cite{pq1,univ}.
    When it is applied to a Kagome lattice \cite{long}, we also find another CDW+VBS states in the easy plane limit $ v< 0, w>0 $, so we believe that
    the CDW-VBS state may be a common and robust state  in any frustrated lattices.
    It is interesting to identify such kind of states by QMC simulations and realize them in near future cold atom experiments.
    The dual relations between  the correlations of bosons and those of vortices explored in this paper should also shed
    lights on exploring duality relations  between correlations functions in other strongly correlated physical systems.

 We thank Fuchun Zhang and Zidan Wang for the hospitality
during our visit at Hong Kong university which initiated the collaboration.
YC was supported by the NSFC-10874032 and 11074043, the State Key
Programs of China (Grant no. 2009CB929204) and Shanghai Municipal
Government. JY was supported by NSF-DMR-1161497, NSFC-11074173,11174210,
Beijing Municipal Commission of Education under grant No.PHR201107121, at
KITP was supported in part by the NSF under grant No. PHY-0551164.
JY thanks Han Pu for his hospitality during his visit at Rice
university where part of this work was done.


\begin{thebibliography}{99}
\bibitem{dof} Such a relation is not needed in the DOF which was
constructed from general symmetry principles in \cite{pq1,tri}.
\end{thebibliography}
\end{document}